\documentclass[12pt]{iopart}
\usepackage{graphicx}
\usepackage{setspace}
\usepackage{texdraw}
\usepackage{amsfonts}
\font\mybbb=msbm10
\def\Bbb#1{\mbox{\mybbb #1}}
\newcommand{\be}{\begin{equation}}
\newcommand{\ee}{\end{equation}}
\newcommand{\bea}{\begin{eqnarray}}
\newcommand{\eea}{\end{eqnarray}}
\newcommand{\eu}{{\rm e}}
\newcommand{\ii}{{\rm i}}

\begin{document}
\title{Renyi Entropy of the XY Spin Chain}

\author{F. Franchini $\star$,  \ A. R. Its\dag \ and V. E. Korepin \maltese }

\address{ $\star$ The Abdus Salam ICTP, Strada Costiera 11, Trieste (TS), 34014, Italy}
\address{ \dag\ Department of Mathematical Sciences, Indiana
University-Purdue University Indianapolis, Indianapolis, IN
46202-3216, USA}
\address{$\maltese$ \ C.N.\ Yang Institute for Theoretical Physics, State
University of New York at Stony Brook, Stony Brook, NY 11794-3840,
USA}

\ead{fabio@ictp.it, itsa@math.iupui.edu, korepin@insti.physics.sunysb.edu}

\begin{abstract}
We consider the one-dimensional \mbox{$\mathrm{XY}$} quantum spin
chain in a transverse magnetic field. We are  interested in the
Renyi entropy of a block of $\mathrm{L}$ neighboring spins at zero
temperature on an infinite lattice. The Renyi entropy is essentially
the trace of some power $\alpha$ of the density matrix of the block.
We calculate the asymptotic for $\mathrm{L}\to \infty$ analytically
in terms of Klein's elliptic $\lambda$ - function.  We study the
limiting entropy as a function of its parameter $\alpha$. We show
that up to the trivial addition terms and multiplicative factors,
and after a proper re-scaling, the Renyi entropy is an automorphic
function with respect to a certain subgroup of the modular group;
moreover, the subgroup depends on whether the magnetic field is
above or below its critical value. Using this fact, we derive the
transformation properties of the Renyi entropy under the map $\alpha
\to \alpha^{-1}$ and show that the entropy becomes an elementary
function of the magnetic field and the anisotropy when
 $\alpha$ is a integer power of $2$, this includes the purity $tr \rho^2$. We also analyze the
behavior of the entropy as $\alpha \to 0$ and $\infty$ and  at  the
critical magnetic field  and in the isotropic limit  [XX model].

\end{abstract}

%\newpage
\section{Introduction}
Entanglement is a resource for quantum control  \cite{ben}. It is necessary for
building quantum computers. Different measures of entanglement are used in the
literature. For pure systems [considered here] the von Neumann entropy of a
subsystem is the most popular measure \cite{vidal, vidal1,cardy, GRAC, keat,
pes}. The subsystem is a large block of spins in the unique ground state of a
spin Hamiltonian. In this paper we evaluate the Renyi entropy of the subsystem.
The Renyi entropy was discovered in information theory \cite{renyi,abe,
BD,hb,L}, it is essentially the trace of a power of the density matrix. For
physics the Renyi entropy is important, because once we know the value of the
trace of every power of the density matrix then we can reconstruct its whole
spectrum.

The physical system we consider is the anisotropic \mbox{$\mathrm{XY}$} model in
a transverse magnetic field and the entropy we are interested in is
the one of a block of $\mathrm{L}$ neighboring spins at zero
temperature and of an infinite system. The Hamiltonian for this
model can be written as
\begin{eqnarray}
H=-\sum_{n=-\infty}^{\infty}
(1+\gamma)\sigma^x_{n}\sigma^x_{n+1}+(1-\gamma)\sigma^y_{n}\sigma^y_{n+1}
+ h\sigma^z_{n} \label{xxh}
\end{eqnarray}
Here $0<\gamma$ is the anisotropy parameter; $\sigma^x_n$,
$\sigma^y_n$ and  $\sigma^z_n$ are the Pauli matrices and $0 \ge h$ is the
magnetic field. The model was solved in
\cite{Lieb,mccoy,mccoy2,gallavotti}.
We are going to calculate the bipartite block entropy of the ground
state $|GS\rangle$ of the system.

\begin{figure}
\begin{center}
   \dimen0=\textwidth
   \advance\dimen0 by -\columnsep
   \divide\dimen0 by 2
\includegraphics[width=\columnwidth]{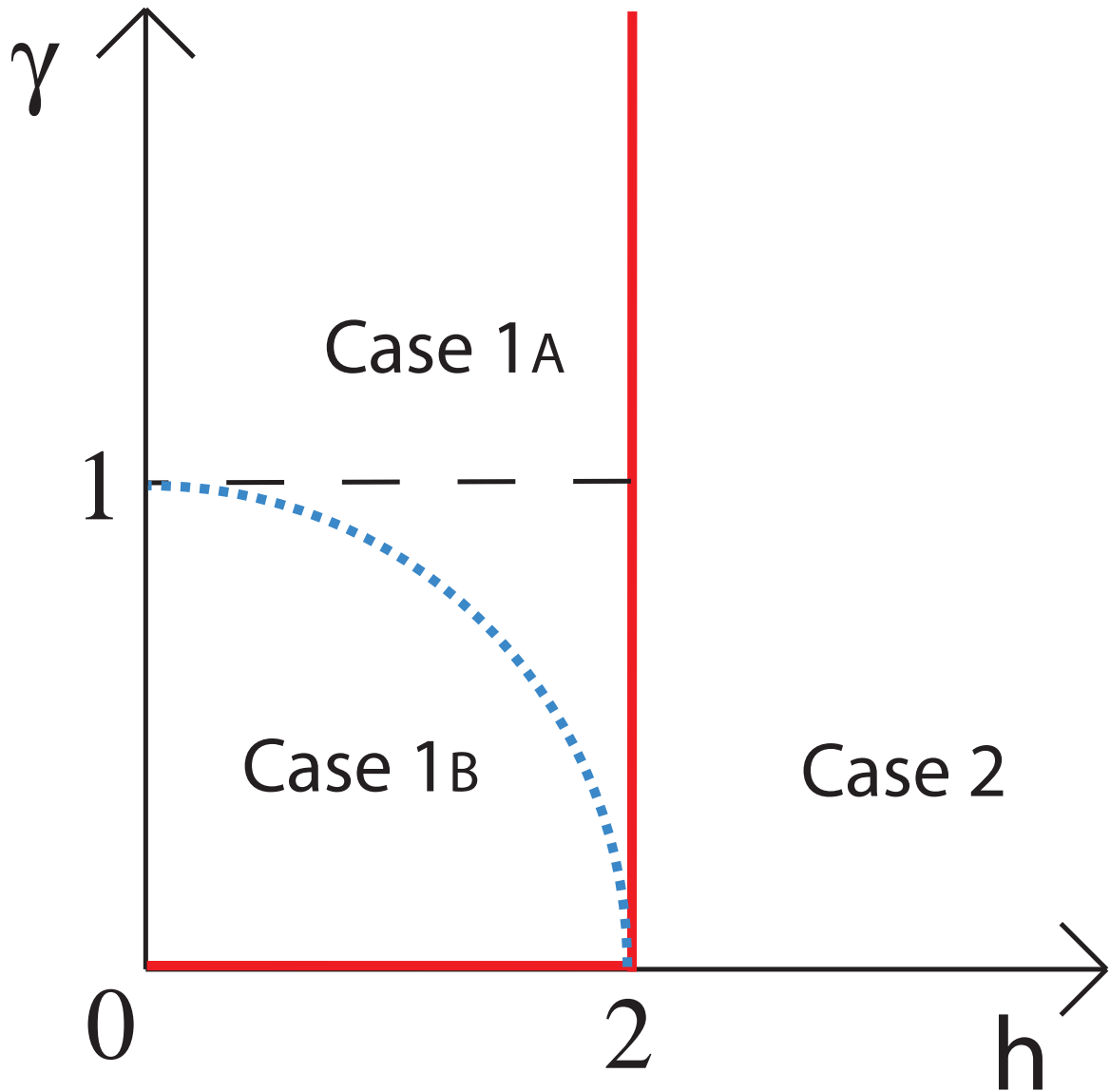}
\caption {Phase diagram of the anisotropic $XY$ model in a
constant magnetic field (only $\gamma \ge 0$ and $h \ge 0$ shown).
The three cases {\sc $2$, $1$a, $1$b}, considered in this paper,
are clearly marked. The critical phases ($\gamma = 0$, $h \le 2$
and $h = 2$) are drawn in bold lines (red, online). The boundary
between cases {\sc $1$a} and {\sc $1$b}, where the ground state is
given by two degenerate product states, is shown as a dotted line
(blue, online). The Ising case ($\gamma = 1$) is also indicated,
as a dashed line.}
   \label{phasediagram}
\end{center}
\end{figure}

The \mbox{$\mathrm{XY}$} model can be mapped exactly into a system of free fermions with
spectrum given by
\begin{equation}
   \epsilon_k = 4 \sqrt{ \left( \cos k - h/2 \right)^2 + \gamma^2 \; \sin^2 k } \; .
\end{equation}
We can read the phase diagram of the model from its spectrum and
identify that it is critical for $\gamma = 0$, $h \le 2$ (corresponding
to the isotropic \mbox{$\mathrm{XY}$} model, or \mbox{$\mathrm{XX}$}
model) and at the critical magnetic field $h = h_c = 2$. For $h = h_f (\gamma) = 2 \sqrt{ 1 - \gamma^2}$
(factorizing field) the ground state can be written as a product
state, as it was found in \cite{shrock}, and is doubly degenerate:
\begin{eqnarray}
   |GS_1\rangle & = & \prod_{n\in \mbox{\rm lattice}}\left[ \cos(\theta)| \uparrow _n \rangle + \sin(\theta)| \downarrow _n \rangle \right] \; ,
  \nonumber   \\
  |GS_2\rangle & = &  \prod_{n\in \mbox{\rm lattice}}\left[ \cos(\theta)| \uparrow _n\rangle - \sin(\theta)| \downarrow _n \rangle \right] \; ,
  \label{deg}
\end{eqnarray}
where  $\cos^2 (2 \theta) =(1-\gamma)/(1+\gamma)$. Off this line, the ground
state of the model $|GS\rangle$ is in continuity with the state
\begin{equation}
   |GS\rangle_{h=h_f (\gamma)} = |GS_1 \rangle + |GS_2 \rangle \; .
   \label{GSfact}
\end{equation}
The line $h = h_f (\gamma)$ is not a phase transition, but the entropy has a
weak singularity across it, since its derivative, although finite, is
discontinuous. In Fig.~\ref{phasediagram} we show the phase diagram of the
\mbox{$\mathrm{XY}$} model and mark the three regions where we calculate the
different expressions of the entropy.

We shall calculate the entropy of a block of $L$ neighboring spins
(a subsystem) of the ground state $|GS\rangle$ as a measure of the entanglement
between this block  and the rest of the chain. We treat the whole
chain as a binary system $|GS\rangle = |A \& B\rangle $. We denote
this block of $L$ neighboring spins by subsystem A and the rest of
the chain by subsystem B. The density matrix of the ground state can
be denoted by \mbox{$\rho_{AB}=|GS\rangle \langle GS|$}. The reduced
density matrix of subsystem A is \mbox{$\rho_A= Tr_B(\rho_{AB})$}.
Then, the von Neumann entropy $S(\rho_A)$ and the R\'enyi entropy
$S_{\alpha}(\rho_A)$ of the block of spins can be evaluated by the
expression
\begin{eqnarray}
S(\rho_A)&=&-Tr(\rho_A \ln\rho_A), \label{edif}\\
S_{\alpha}(\rho_A)&=&\frac{1}{1-\alpha}\ln Tr(\rho_A^{\alpha}),
\qquad \alpha\neq 1 ~~\textrm{and}~~ \alpha>0 .\label{olds}
\end{eqnarray}
Here the power $\alpha$ is a parameter. When evaluated for 1-dimensional
critical theories, these entropies diverge logarithmically with the size of the
block, while they saturate to a constant in the presence of a gap
\cite{arealaw}.

For the isotropic version of the XY model $\gamma=0$ we evaluated R\'enyi
entropy of a large block of spins in \cite{jin}. The von Neumann entropy of the
block in the XY model was calculated in \cite{ijk1,ijk2,fijk1,fijk2}. The
methods of Toeplitz determinants \cite{wu,fisher,basor,tracy,abbs,aban}, as
well as the techniques based on integrable Fredholm
operators\cite{iiks,sla,korepin}, have been used for the evaluation of the von
Neumann entropy of this model \cite{jin,jpaits}.

In this paper we  evaluate the R\'enyi entropy, which is the natural
generalization of the von Neumann entropy \cite{renyi}. When $\alpha
\to 1$, the  R\'enyi entropy turns into the von Neumann entropy.

\section{Renyi Entropy.}

The von Neumann Entropy of the block of spins has been calculated in
\cite{jpaits} and  \cite{pes}. We shall use the same notations and
introduce an elliptic parameter:
\begin{eqnarray} \label{kmain}
k=\left\{\begin{array}{l} \sqrt{(h/2)^2+\gamma^2-1}\; /\; \gamma ,
\;\;\;\;\;\;\;\;\;\;\;\;\;\;\;\;\mbox{Case
1a:~$4(1-\gamma^2)<h^2<4$;} \\ [0.3cm]
\sqrt{({1-h^2/4-\gamma^2})/({1-h^2/4})}\; ,\;\;\;\mbox{Case 1b:~$h^2<4(1-\gamma^2)$;} \\
[0.3cm]
       \gamma\; / \;\sqrt{(h/2)^2+\gamma^2-1} ,\;\;\;\;\;\;\;\;\;\;\;\;\;\;\;\;\mbox{Case 2~:~$h>2$.}
       \end{array}\right.
\end{eqnarray}
We shall also use  the complete elliptic integral of the first
kind
\begin{equation}\label{ellint}
I(k) = \int_{0}^{1}\frac{dx}{\sqrt{(1-x^2)(1 - k^{2}x^{2})}}
\end{equation}
and the modulus
\begin{equation}
   \tau_0= I(k')/I(k), \qquad \qquad k'=\sqrt{1-k^2},
\end{equation}
%\begin{equation}
%  \lambda_{m} =
%  \tanh \left(m + \frac{1-\sigma}{2}\right)\pi \tau_{0},
%  \label{zerosMay}
%\end{equation}
%here $\sigma = 1$ for $h<2$ and $\sigma = 0$ for $h>2$.
as well as:
\be
   \epsilon \equiv \pi \tau_0, \qquad   q \equiv \eu^{-\epsilon} = \eu^{-\pi I(k')/I(k)}.
\ee

We will need the following identities as well \cite{WW}:
\be
\label{July1}
 \prod_{m=0}^\infty \left( 1 + q^{2m+1} \right) =
   \left( {16 q \over k^2 k'^2} \right)^{1/24}
\ee
\be
   \prod_{m=1}^\infty \left( 1 + q^{2m} \right) =
   \left( {k^2 \over 16 q k'} \right)^{1/12}.
   \label{July3}
\ee
%This equations is a direct consequence of the classical infinite
%product and Fourier series representations for the Jacobi elliptic
%functions \cite{WW}.

Now let us start the evaluation of the {\bf {Renyi entropy}} of a
block of $L$ neighboring spins. It  can be represented \cite{jin} as
\be \label{renyiLdef}
 S_R (\rho_A,\alpha) = {1 \over 1- \alpha} \sum_{k=1}^L
   \ln \left[ \left( {1 + \nu_k \over 2 } \right)^\alpha
   + \left( {1 - \nu_k \over 2 } \right)^\alpha \right] ,
\ee
where the numbers
$$
\pm i\nu_{k}, \quad k = 1, ..., L
$$
are the eigenvalues of a certain block Toeplitz matrix.
In \cite{jpaits} it is shown that in the large $L$ limit the
eigenvalues $\nu_{2m}$ and $\nu_{2m+1}$ merge to
the number $\lambda_{m}$ defined below in eq  (\ref{lam}),
$$
\nu_{2m}, \nu_{2m+1} \to \lambda_{m}.
$$
Hence, the Renyi entropy in the large $L$ limit can be identified
with the convergent series,
\be \label{rendef0}
   S_R (\rho_A,\alpha) = {1 \over 1-\alpha} \sum_{m=-\infty}^\infty
   \ln \left[ \left( {1 + \lambda_m \over 2 } \right)^\alpha
   + \left( {1 - \lambda_m \over 2 } \right)^\alpha \right] ,
\ee with
\begin{equation}
  \lambda_{m} =
  \tanh \left(m + \frac{1-\sigma}{2}\right)\pi \tau_{0}. \label{lam}
\end{equation}
The summation of the series can be done following the same approach
as in the case of the von Neuman entropy (cf. \cite{pes}).

\subsection{$h>2$}

With $\epsilon \equiv \pi \tau_0$, we have \bea
   1 + \lambda_m & = & 2 {1 \over 1 + \eu^{-(1+2m)\epsilon} } \\
   1 - \lambda_m & = & 2 { \eu^{-(1+2m)\epsilon} \over 1 + \eu^{-(1+2m)\epsilon}
   }.
\eea Then, the entropy is: \bea
    S_R & = & {1 \over 1 - \alpha } \sum_{m=-\infty}^\infty
    \ln \left[ \left( {1 \over 1 + \eu^{-(1+2m)\epsilon}} \right)^\alpha +
    \left( {\eu^{-(1+2m)\epsilon} \over 1 + \eu^{-(1+2m)\epsilon}} \right)^\alpha \right]
    \nonumber \\
    & = & {2 \over 1 - \alpha } \sum_{m=0}^\infty
    \ln \left[ \left( {1 \over 1 + \eu^{-(1+2m)\epsilon}} \right)^\alpha +
    \left( {\eu^{-(1+2m)\epsilon} \over 1 + \eu^{-(1+2m)\epsilon}} \right)^\alpha \right]
    \nonumber \\
    & = & {2 \over 1 - \alpha } \sum_{m=0}^\infty
    \ln \left[ {1 + \eu^{-\alpha (1+2m) \epsilon} \over \left( 1 + \eu^{-(1+2m)\epsilon} \right)^\alpha } \right]
    \nonumber \\
    & = & {2 \over 1 - \alpha } \sum_{m=0}^\infty
    \ln \left[ 1 + \eu^{- \alpha (1+2m) \epsilon} \right] - {2 \alpha \over 1 - \alpha }
     \sum_{m=0}^\infty\ln
    \left[ 1+ \eu^{-(1+2m)\epsilon}  \right] .
    \label{renyi1}
\eea Summing the second term is straightforward, using (\ref{July1}):
\bea
   \hskip -1cm
   - {2 \alpha \over 1 - \alpha } \sum_{m=0}^\infty \ln
    \left[ 1+ \eu^{-(1+2m)\epsilon}  \right]  & = &
   - {2 \alpha \over 1 - \alpha } \ln \prod_{m=0}^\infty
   \left( 1+ q^{2m+1} \right) \nonumber \\
   & = & -{1 \over 12} { \alpha \over 1 - \alpha } \; \left[
   \ln q +
   \ln \left( {16 \over k^2 k'^2 } \right) \right] \label{first},
\eea where, as usual,
\be \label{qk}
   q \equiv \eu^{-\pi I(k')/I(k)}.
\ee

In order to sum up the first term we notice that identity (\ref{July1}) can be
interpreted as the evaluation of the product in the left hand side in terms of
the function $k \equiv k(q)$ defined implicitly by equation (\ref{qk}). A
fundamental fact of the theory of elliptic functions is that the function
$k(q)$  admits an {\it explicit} representation in terms of the
theta-constants. Indeed, the following formulae take place (see e.g.
\cite{WW}): \be
   \hskip -1cm
   k(q) = {\theta_2^2 (0, q) \over \theta_3^2 (0, q) } \; , \qquad \qquad
   k'(q) = {\theta_4^2 (0, q) \over \theta_3^2 (0, q) } \; ,
  \label{thetaks1}
\ee where $\theta_{j}(z|q)$, $j = 1,2,3,4$ are the Jacobi theta-functions. We
remind (see  again \cite{WW}) that  the theta functions are defined for any
$|q| < 1$ by the following Fourier series \be \label{t1} \theta_{1}(z, q) =
i\sum_{n=-\infty}^{\infty}(-1)^n q^{\left(\frac{2n-1}{2}\right)^2}e^{2
iz\left(n-\frac{1}{2}\right)}\; , \ee \be \label{t2} \theta_{2}(z, q) =
\sum_{n=-\infty}^{\infty} q^{\left(\frac{2n-1}{2}\right)^2}e^{2
iz\left(n-\frac{1}{2}\right)}\; , \ee \be \label{t3} \theta_{3}(z, q) =
\sum_{n=-\infty}^{\infty} q^{n^2}e^{2 iz n}\; , \ee \be \label{t4}
\theta_{4}(z, q) = \sum_{n=-\infty}^{\infty}(-1)^n q^{n^2}e^{2 iz n}\;.
\ee
In
particular, it follows that the functions
\be
   \label{analk}
   k'(q) \quad \mbox{and}\quad q^{-1/2}k(q) \;,
\ee are analytic on the unit disc $|q| < 1$. It is also worth mentioning the
classical formula for the integral $I(k)$,
\be
   I(k) = \frac{\pi}{2}\theta_{3}^{2}(0, q).
\ee

Put now,
\be \label{kalphadef}
k_{\alpha}: = k(q^{\alpha})\; ,
\ee
where $q$ is the $q$ - parameter corresponding via (\ref{qk})
to the original elliptic parameter $k$ from (\ref{kmain}).
Then for the first term in (\ref{renyi1}) we will have
\bea
   \hskip -1cm
    {2  \over 1 - \alpha } \sum_{m=0}^\infty \ln
    \left[ 1+ \eu^{-\alpha(1+2m)\epsilon}  \right]  & = &
    {2  \over 1 - \alpha } \ln \prod_{m=0}^\infty
   \left( 1+( q^{\alpha})^{2m+1} \right) \nonumber \\
   & = & {1 \over 12} { 1\over 1 - \alpha } \; \left[
  \alpha \ln q +
   \ln \left( {16 \over k^2_{\alpha} k'^2_{\alpha} } \right) \right] \label{second}.
\eea
Substituting  this expression together with  (\ref{first})  into (\ref{renyi1}),
we arrive at the equation,
$$
S_R = \frac{1}{12}\frac{1}{1 - \alpha}
\ln\left(\frac{16}{k^2_{\alpha}k'^2_{\alpha}}\right)
-  \frac{1}{12}\frac{\alpha}{1 - \alpha}
\ln\left(\frac{16}{k^2 k'^2}\right)
$$
\be \label{renfinal1}
=   \frac{1}{6}\frac{\alpha}{1 - \alpha}\ln (kk')
- \frac{1}{6}\frac{1}{1 - \alpha}
\ln (k_{\alpha}k'_{\alpha})  + {1 \over 3} \ln 2,
\ee
which in turns yields the following final expression for the
Renyi entropy.
\be\label{renfinal2}
   \hskip -1.5cm
   S_R(\rho_{A},\alpha) = {1 \over 6} \; { \alpha \over 1 - \alpha} \; \ln \left( k \; k' \right)
   - {1 \over 3} \; { 1 \over 1 - \alpha} \; \ln \left(
   { \theta_2 (0, q^\alpha) \; \theta_4 (0, q^\alpha) \over \theta_3^2 (0, q^\alpha)} \right) \;
    + {1 \over 3} \ln 2.
\ee
Here, the elliptic parameter $k$ is defined in (\ref{kmain}),
$k' = \sqrt{1-k^2}$, the modulus parameter $q$ is given by equation
(\ref{qk}) where $I(k)$ is the complete elliptic integral (\ref{ellint}),
and the theta functions $\theta_j(z, q)$ are defined by the series
(\ref{t1} - \ref{t4}).

\subsection{$h<2$}

In this case we have
\be
   \lambda_m = \tanh ( m \pi \tau_0 ) =
   {\eu^{2m \epsilon} - 1 \over \eu^{2m \epsilon} + 1},
\ee
where, as usual,
\be
   \epsilon \equiv \pi \tau_0.
\ee

The entropy is:
\bea
    S_R & = & {1 \over 1 - \alpha } \sum_{m=-\infty}^\infty
    \ln \left[ \left( {1 \over 1 + \eu^{-2m \epsilon}} \right)^\alpha +
    \left( {\eu^{-2m \epsilon} \over 1 + \eu^{-2m \epsilon}} \right)^\alpha \right]
    \nonumber \\
    & = & {2 \over 1 - \alpha } \sum_{m=1}^\infty
    \ln \left[ \left( {1 \over 1 + \eu^{-2m \epsilon}} \right)^\alpha +
    \left( {\eu^{-2m \epsilon} \over 1 + \eu^{-2m \epsilon}} \right)^\alpha \right] +
    {1 \over 1 - \alpha} \ln \left( 2 {1 \over 2^\alpha} \right)
    \nonumber \\
    & = & {2 \over 1 - \alpha } \sum_{m=1}^\infty
    \ln \left[ 1 + \eu^{- 2 \alpha m \epsilon} \right] - {2\alpha \over \alpha - 1}
     \sum_{m=1}^\infty  \ln
    \left[ 1+ \eu^{-2m \epsilon}  \right]  + \ln 2.
    \label{renyi3}
\eea Again the second term can be immediately summed using
(\ref{July3}):
\bea
    - {2 \alpha \over 1 - \alpha } \sum_{m=1}^\infty \ln
    \left[ 1+ \eu^{-2m \epsilon}  \right] & = &
    - {2 \alpha \over 1 - \alpha } \ln \prod_{m=1}^\infty
    \left( 1+ q^{2m} \right) \nonumber \\
    & = & - {1 \over 6} {\alpha \over 1 - \alpha} \left[
    \ln \left( { k^2 \over 16 k'} \right) -\ln q \right],\label{case21}
\eea
where, as usual,
\be
   q \equiv \eu^{-\pi I(k')/I(k)}.
\ee
The first term, as in the previous case, admits the similar
representation involving the elliptic parameter $k_{\alpha}
\equiv k(q^{\alpha})$,
\be\label{case22}
    {2 \over 1 - \alpha } \sum_{m=1}^\infty \ln
    \left[ 1+ \eu^{-2 \alpha m \epsilon}  \right] =
    {1 \over 6} {1 \over \alpha - 1 } \left[
    \ln \left( { k^2_{\alpha} \over 16 k'_{\alpha}} \right) - \alpha \ln q \right].
\ee Using (\ref{case21}) and (\ref{case22}) in (\ref{renyi3}), we obtained that
\begin{eqnarray}
   S_R & = & \frac{1}{6}\frac{1}{1 - \alpha}
\ln\left(\frac{k^2_{\alpha}}{16k'_{\alpha}}\right)
-  \frac{1}{6}\frac{\alpha}{1 - \alpha}
\ln\left(\frac{k^2}{16 k'}\right) + \ln 2
\nonumber \\
\label{renfinal3}
 & = &   \frac{1}{6}\frac{\alpha}{1 - \alpha }\ln\left(\frac{k'}{k^2}\right)
+ \frac{1}{6}\frac{1}{1 - \alpha}
\ln \left(\frac{k_{\alpha}^2}{k'_{\alpha}}\right)  + {1 \over 3} \ln 2,
\end{eqnarray}
which in turns yields the following final expression for the
Renyi entropy in the case $h < 2$.
\be\label{renfinal4}
   \hskip -1.5cm
   S_R(\rho_{A},\alpha) = {1 \over 6} \; {\alpha \over 1 - \alpha } \; \ln
   \left( {k'\over k^2 } \right) + {1 \over 3} \; {1 \over 1 - \alpha} \;
   \ln \left( { \theta_2^2 (0, q^\alpha) \over \theta_3 (0, q^\alpha) \; \theta_4 (0, q^\alpha)  } \right)
   + {1 \over 3} \ln 2.
\ee
Here, as before, the elliptic parameter $k$ is defined in (\ref{kmain}),
$k' = \sqrt{1-k^2}$, the modulus parameter $q$ is given by equation
(\ref{qk}) where $I(k)$ is the complete elliptic integral (\ref{ellint}),
and the theta functions $\theta_j(z|q)$ are defined by the series
(\ref{t1} - \ref{t4}).

{\bf Remark.} One can wonder about an apparent tautological character
of the formulae (\ref{renfinal2}) and (\ref{renfinal4}). Indeed, they seem just
to re-express one $q$-series ($S_R(\rho_{A},\alpha)$) in terms of the another
($\theta_{j}(0 | q)$). The important  point however is that the $q$-series
representing the theta-constants place the object of interest, i.e. the Renyi
entropy, in the well - developed realm of classical elliptic functions.
In fact, to solve a problem in terms of Jacobi
theta- function is as good as to solve it in terms of, say,
elementary exponential function (after all, the exponential function
is also an infinite series !). The crucial thing is that
a lot is known about the properties of the theta-constants and this allows
a quite comprehensive study of the Renyi entropy both numerically
and analytically. In the next section we will demonstrate the efficiency
of equations (\ref{renfinal2}) and (\ref{renfinal4}).

\section{Renyi Entropy. The Analysis.}

When studying the analytic properties of the Renyi entropy with
respect to the variable $\alpha$, it is convenient to pass from the
modulus parameter $q$ to the (more standard) modulus parameter
$\tau$ defined by the relations, \be \label{taudef} q = e^{\pi i
\tau}, \quad  \tau= i\frac{I(k')}{I(k)} \equiv i\tau_{0}, \quad \Im
\tau > 0. \ee The theta functions $\theta_{j}(z, q)$ then become the
functions, \be \label{thetatau} \theta_{j}(z|\tau):= \theta_{j}(z,
e^{\pi i \tau}),\quad j = 1, 2, 3, 4, \ee which are holomorphic for
all $z$ and for all $\tau$ from the upper half plane, \be
\label{domaintau} \Im \tau > 0. \ee Using these new notations, the
above obtain formulae for the Renyi entropy can be rewritten as
\be\label{renfinal2new}
   \hskip -2cm
   S_R(\rho_{A},\alpha) = {1 \over 6} \; { \alpha \over 1 - \alpha} \; \ln \left( k \; k' \right)
   - {1 \over 3} \; { 1 \over 1 - \alpha} \; \ln \left(
   { \theta_2 (0| \alpha i \tau_{0}) \; \theta_4 (0| \alpha i \tau_{0})
   \over \theta_3^2 (0| \alpha i \tau_{0})} \right)\;
 + {1 \over 3} \ln 2,
\ee
for $h> 2$ and
\be\label{renfinal4new}
   \hskip -2cm
   S_R(\rho_{A},\alpha) = {1 \over 6} \; {\alpha \over 1 - \alpha } \; \ln
   \left( {k'\over k^2 } \right) + {1 \over 3} \; {1 \over 1 - \alpha} \;
   \ln \left( { \theta_2^2 (0| \alpha i\tau_{0})
   \over \theta_3 (0| \alpha i \tau_{0}) \; \theta_4 (0| \alpha i \tau_{0})  } \right)
   + {1 \over 3} \ln 2,
\ee for $h < 2$. To proceed with the analysis of these expressions
as functions of $\alpha$, we will need some pieces of the general
theory of Jacobi functions $\theta_{j}(z|\tau)$ which we collect in
Appendix A.

Our first observation is that the domain of analyticity (\ref{domaintau}) and
the positiveness of the parameter $\tau_{0}$ indicate that the all three
theta-constants, i.e. $\theta_{2}(0|\alpha i\tau_{0})$, $\theta_{3}(0|\alpha
i\tau_{0})$, and $\theta_{4}(0|\alpha i\tau_{0})$ are analytic in the right
half plane of the complex $\alpha$ - plane: \be \label{domainalpha} \Re \alpha
> 0. \ee Simultaneously we notice that inequality (\ref{nonzero}) implies that
the theta-ratios appearing in the right hand sides of (\ref{renfinal2new}) and
(\ref{renfinal4new}) are never zero. Hence we can claim that the Renyi entropy,
as a function of $\alpha$, is analytic in the right half plane
(\ref{domainalpha}), with the possible pole at $\alpha = 1$. However, since as
$\alpha \to 1$ the theta-ratios in (\ref{renfinal2new}) and
(\ref{renfinal4new}) become the square roots of the product $kk'$ and of the
ratio $k^2/k'$, respectively (see also (\ref{renfinal1}) and
(\ref{renfinal3})), the singularity at $\alpha = 1$ is, in fact, removable and
we can write that
$$
S_{R}(\rho_{A}, 1) = -\frac{1}{6}\ln(kk') +\frac{1}{3}\ln2
+\frac{1}{3}\frac{d}{d\alpha}
\ln \left(
   { \theta_2 (0| \alpha i \tau_{0}) \; \theta_4 (0| \alpha i \tau_{0})
   \over \theta_3^2 (0| \alpha i \tau_{0})} \right)_{\alpha = 1}
$$
\be \label{ren_neu1}
\equiv
-\frac{1}{6}\ln(kk') +\frac{1}{3}\ln2
+\frac{1}{6}\frac{d}{d\alpha}
\ln(k_{\alpha}k'_{\alpha})|_{\alpha = 1}
\ee
for $h >2$ and
$$
S_{R}(\rho_{A}, 1) = -\frac{1}{6}\ln\left(\frac{k'}{k^2}\right) +\frac{1}{3}\ln2
+\frac{1}{3}\frac{d}{d\alpha}
\ln \left(
   { \theta_3 (0| \alpha i \tau_{0}) \; \theta_4 (0| \alpha i \tau_{0})
   \over \theta_2^2 (0| \alpha i \tau_{0})} \right)_{\alpha = 1}
$$
\be \label{ren_neu2}
\equiv
-\frac{1}{6}\ln\left(\frac{k'}{k^2}\right) +\frac{1}{3}\ln2
+\frac{1}{6}\frac{d}{d\alpha}
\ln\left(\frac{k'_{\alpha}}{k^2_{\alpha}}\right)|_{\alpha = 1}
\ee
for $h <2$. It is an exercise in the theory of elliptic functions
to show that the expressions on the right hand sides of (\ref{ren_neu1})
and  (\ref{ren_neu2}) are in fact the respective Von Neumann entropies
calculated in \cite{jin,jpaits,pes}:
\begin{eqnarray}
   {\hskip -1cm } S(\rho_A) =  \frac {1} {6} \left [\;\ln{ \frac {4} {k \; k'}} + (k^2-k'^2)
   \frac {2 I(k) I(k')} {\pi} \right ] \; ,
   & \qquad  & h > 2 \; ; \\
   {\hskip -1cm }  S(\rho_A) =   \frac {1} {6} \left [\;\ln{ \left (\frac { 4 k^2} { k'}\right )} +
   \left( 2 - k^2 \right)  \frac {2 I(k) I(k')} {\pi} \right ] \; ,
   & & h < 2 \; .
\end{eqnarray}

This fact, i.e. the statement that
\begin{equation}
  \lim_{\alpha \to 1}S_{R}(\rho_{A}, \alpha) = S (\rho_{A}) \; ,
\end{equation}
can be of course obtained via much more elementary
calculations based on the original series representation
(\ref{rendef0}) for $S_{R}(\rho_{A}, \alpha)$.

Consider now the two other critical cases: $\alpha \to 0$ and $\alpha \to \infty$.

\subsection{$\alpha \to \infty$}
The limit of large $\alpha $ is interesting for the single copy entanglement
suggested by M. Plenio and J. Eisert\cite{EisCram}. In fact, the Renyi entropy
contains information about all eigenvalues of the density matrix and we can
extract the largest eigenvalue [maximum probability $p_M$] from the limit
$\alpha \to \infty$ ($S_\alpha (\rho_A) \to - \ln p_M$).

Using the first series from equations (\ref{jacobi2}) -
(\ref{jacobi4}) we obtain at once that \be \label{infin1}
\frac{\theta_{2}(0|\alpha i \tau_{0})\theta_{4}(0|\alpha i
\tau_{0})} {\theta^{2}_{3}(0|\alpha i \tau_{0})} = 2 e^{-\frac{\pi
\alpha \tau_{0}}{4}} \left( 1 + O\left(e^{-\alpha \pi
\tau_{0}}\right)\right) \ee and \be \label{infin2}
\frac{\theta^{2}_{2}(0|\alpha i \tau_{0})} {\theta_{3}(0|\alpha i
\tau_{0})\theta_{4}(0|\alpha i \tau_{0})} = 4 e^{-\frac{\pi \alpha
\tau_{0}}{2}} \left( 1 + O\left(e^{-2\alpha \pi
\tau_{0}}\right)\right), \ee as $\alpha \to \infty$, $-\pi/2 < \arg
\alpha < \pi/2$. Plugging these estimates in (\ref{renfinal2new})
and (\ref{renfinal4new}) and recalling that $\tau_{0} = I(k')/I(k)$,
we arrive at the following description of the Renyi entropy in the
large $\alpha$ limit. \be \label{largealpha1} S_{R}(\rho_{A},\alpha)
= \frac{\alpha}{1 - \alpha} \left(\frac{1}{6}\ln\frac{kk'}{4} +
\frac{\pi}{12}\frac{I(k')}{I(k)}\right) \ee
$$
 +
O\left(\frac{1}{\alpha}e^{-\alpha \pi \tau_{0}}\right)
$$
$$
= - \frac{1}{6}\ln\frac{kk'}{4} +
\frac{\pi}{12}\frac{I(k')}{I(k)} +
O\left(\frac{1}{\alpha}\right),
$$
$$
\alpha \to \infty,\quad -\frac{\pi}{2} < \arg \alpha < \frac{\pi}{2},
$$
for $h>2$, and
\be \label{largealpha2}
S_{R}(\rho_{A},\alpha) = \frac{\alpha}{1 - \alpha}
\left(\frac{1}{6}\ln\frac{k'}{4k^2} -
\frac{\pi}{6}\frac{I(k')}{I(k)} \right) + \frac{1}{1 - \alpha}\ln 2
\ee
$$
 +
O\left(\frac{1}{\alpha}e^{-2\alpha \pi \tau_{0}}\right)
$$
$$
= -\frac{1}{6}\ln\frac{k'}{4k^2} +
\frac{\pi}{6}\frac{I(k')}{I(k)}
+ O\left(\frac{1}{\alpha}\right),
$$
$$
\alpha \to \infty,\quad -\frac{\pi}{2} < \arg \alpha < \frac{\pi}{2},
$$
for $h<2$.
Alternatively, these estimates can be easily extracted
from the original series representations, i.e. equations
(\ref{renyi1}) and (\ref{renyi3}), with the help of the
identities (\ref{July1}) and (\ref{July3}). In other words,
the theta-summation of the series (\ref{renyi1}) and (\ref{renyi3})
is not really needed for the large values of the parameter $\alpha$.

{\it Remark.} The asymptotic representations (\ref{largealpha1}) and
(\ref{largealpha2}) are only valid for the bulk of the $XY$ model, i.e. away
from critical lines $\gamma=0$ or $h=2$. Near the critical points, when $\gamma
\neq 0$ and $h \to 2$, or $ \gamma \to 0$ and $h <2$, the module parameter
$\tau_{0}$ becomes small and the estimates (\ref{largealpha1}) and
(\ref{largealpha2}) are not valid unless the double scaling  condition,
\be
   \alpha \tau_{0} \to \infty
\ee takes place.

\subsection{$\alpha \to 0$}

This is where the theta-formulae help. Indeed, using the second
series from the Jacobi identities (\ref{jacobi2}) - (\ref{jacobi4}),
we arrive at the estimates,

\be
   \label{zero1}
   \frac{\theta_{2}(0|\alpha i \tau_{0})\theta_{4}(0|\alpha i
   \tau_{0})} {\theta^{2}_{3}(0|\alpha i \tau_{0})} = 2 e^{-\frac{\pi}{4 \alpha
   \tau_{0}}} \left( 1 + O\left(e^{-\frac{\pi}{\alpha \tau_{0}}}\right)\right)
\ee and
\be
   \label{zero2}
   \frac{\theta^2_{2}(0|\alpha i \tau_{0})}
   {\theta_{3}(0|\alpha i \tau_{0})\theta_{4}(0|\alpha i \tau_{0})} = \frac{1}{2}
   e^{\frac{\pi}{4 \alpha \tau_{0}}} \left( 1 + O\left(e^{-\frac{\pi}{ \alpha
   \tau_{0}}}\right)\right)
\ee as $\alpha \: \tau_{0} \to 0$, $-\pi/2 < \arg \alpha < \pi/2$. These
formulae indicate the appearance of a singularity of order $\alpha^{-1}$ in the
Renyi entropy as $\alpha \to 0$. In fact, since we consider the limit of a
large block of spins, the dimension of the corresponding Hilbert space also
goes to infinity. This is the reason for which the Renyi entropy has a
singularity at $\alpha = 0$.

Substituting (\ref{zero1}) and (\ref{zero2}) into (\ref{renfinal2new}) and
(\ref{renfinal4new}), respectively, we obtain the following description of the
Renyi entropy in the small $\alpha$ limit. \be \label{smallalpha1}
S_{R}(\rho_{A},\alpha) = \frac{1}{\alpha(1-\alpha)}
\frac{\pi}{12}\frac{I(k)}{I(k')} + \frac{\alpha}{1-\alpha}\frac{1}{6}\ln
\frac{kk'}{4} + O\left(e^{-\frac{\pi}{\alpha\tau_{0}}}\right) \ee \be
\label{smallalpha11} = \frac{1+\alpha}{\alpha}\frac{\pi}{12}\frac{I(k)}{I(k')}
+ O(\alpha), \ee
$$
\alpha \to 0,\quad -\frac{\pi}{2} < \arg \alpha < \frac{\pi}{2},
$$
for $h>2$, and
\be \label{smallalpha2}
S_{R}(\rho_{A},\alpha) = \frac{1}{\alpha(1-\alpha)}
\frac{\pi}{12}\frac{I(k)}{I(k')} +
\frac{\alpha}{1-\alpha}\frac{1}{6}\ln \frac{k'}{4k^2}
+ O\left(e^{-\frac{\pi}{\alpha\tau_{0}}}\right)
\ee
\be \label{smallalpha3}
= \frac{1+\alpha}{\alpha}\frac{\pi}{12}\frac{I(k)}{I(k')}
+ O(\alpha),
\ee
$$
\alpha \to 0,\quad -\frac{\pi}{2} < \arg \alpha < \frac{\pi}{2},
$$
for $h<2$.

Similar to the case of the Von Neumann entropy dealt with in
\cite{jpaits}, equations (\ref{smallalpha1}) and (\ref{smallalpha2})
can be also used for the evaluation of the small $\tau_{0} \equiv
I(k')/I(k)$ limit of the Renyi entropy with the fixed $\alpha >0$.
This limit (cf. \cite{jpaits}) appears either in the case of the
critical magnetic field, i.e. $\gamma \neq 0$ and $h \to 2$, or when
approaching the $XX$ model, i.e. $\gamma \to 0$ and $h <2$. We shall
now consider these limits.

\subsection{Critical magnetic field: $\gamma \neq 0$ and $ h \to 2$}
This is included in  Case 1a and Case 2 which means that,
\be
 \hskip -1cm
k = 1 -\frac{1}{2\gamma^2}|h-2| + O(|h-2|^2),
\quad k' = \frac{1}{\gamma}|h-2|^2\left(1 +  O(|h-2|)\right),
\ee
and, in turn,
\be \label{critmag}
\pi\frac{I(k)}{I(k')}
= -\ln|2-h| +2\ln 4\gamma + O(|h-2|\ln^2|h-2|),
\ee
$$
 \quad h \to 2, \quad \gamma \neq 0.
$$
This means that in this limit $\tau_0 \to 0$ and we can use (\ref{smallalpha1})
to arrive at the following estimates for the Renyi entropy in the case of the
critical magnetic field,
\be \label{crtimag2} S_{R}(\rho_{A},\alpha) =
\frac{1+\alpha}{\alpha}\Bigl(-\frac{1}{12}\ln|2-h| + \frac{1}{6}\ln 4\gamma
\Bigr) \ee
$$
+ O(|h-2|\ln^2|h-2|).
$$

We notice that the singularity of the Renyi entropy is logarithmic like for the
Von Neumann entropy, but coefficient in front of the logarithm is different and
$\alpha$-dependent.

\subsection{An approach to $XX$ model: $\gamma \to 0$ and $ h < 2$}
This is included in  Case 1b  which means that,
\be
k = 1 -\frac{2\gamma^2}{4-h^2} + O(\gamma^4),
\quad k' = \frac{2\gamma}{\sqrt{4-h^2}}\left(1 +  O(\gamma^2)\right),
\ee
and, in turn,
\be \label{XX}
\pi\frac{I(k)}{I(k')}
= -2\ln \gamma +\ln(4-h^2) +2\ln 2 + O(\gamma \ln^2\gamma),
\ee
$$
\gamma \to 0, \quad h < 2\sqrt{1-\gamma^2}.
$$
Again, since $\tau_0 \to 0$, we can substitute these into (\ref{smallalpha2})
and arrive at the following estimates for the Renyi intropy in the case of the
XX model limit \be \label{XX2} S_{R}(\rho_{A},\alpha) =
\frac{1+\alpha}{\alpha}\Bigl(-\frac{1}{6}\ln\gamma + \frac{1}{12}\ln (4-h^2)
+\frac{1}{6}\ln2 \Bigr) \ee
$$
+ O(\gamma\ln^2\gamma).
$$
We note that if $\alpha =1$ then equations (\ref{crtimag2}) and (\ref{XX2}) transforms to
the respective formulae for the Neumann entropy obtained earlier
in \cite{jpaits}.

\subsection{The factorizing field}

We already showed in the introduction that for $h = h_f (\gamma)
= 2 \sqrt{ 1 - \gamma^2}$ the ground state can be written as
\be
   | GS \rangle = | GS_1 \rangle + | GS_2 \rangle \; ,
   \label{GSfact1}
\ee
where $|GS_{1,2} \rangle$ are the product states given in (\ref{deg})
and clearly have no entropy/entanglement by themselves.

We can calculate the Renyi entropy of the ground state at the factorizing
field by considering the limit $k \to 0$ of (\ref{renfinal4new}).
Remembering that, using (\ref{jacobi2}-\ref{jacobi4}) in this limit
\begin{eqnarray}
   \theta_2 (0 | \alpha \ii \tau \sim 0 ) & \sim &
   2 \left( {k \over 4} \right)^{\alpha /2} \; , \\
   \theta_3 (0 | \alpha \ii \tau \sim 0 ) & \sim &
   \theta_4 (0 | \alpha \ii \tau \sim 0 ) \sim 1 \; ,
\end{eqnarray}
it is easy to show that
\begin{equation}
   S_{R}(\rho_{A},\alpha) = \ln 2
   \label{degS}
\end{equation}
regardless the value of $\alpha$. This result is not surprising and was to be
expected in light of (\ref{GSfact1}). In fact, the limiting density matrix of
the block of spins at the factorizing field is $(1/2) \times I_2$, where $I_2$
is the $2 \times 2$ Identical matrix.

Please note the importance of the order of limits around the factorizing field.
In fact, the expression in (\ref{degS}) is independent of $\alpha$ and
therefore regular in the limit $\alpha \to 0$, while off the factorizing field
line the entropy diverges like in (\ref{smallalpha3}) for $\alpha \to 0$. As
one approaches the factorizing field, $k \to 0$ and therefore $\tau_o \to
\infty$ in such a way that $\alpha \tau_0$ stays constant.

\section{Renyi Entropy and the Modular Functions.}

The square of the elliptic parameter $k$, considered as a function of the
modulus $\tau$, is usually dented as $\lambda(\tau)$, and it is called the {\it
elliptic lambda function } or {\it $\lambda$ - modular function}. We note that
(cf. (\ref{thetaks1})) \be \label{kappadef} \lambda(\tau) =
\frac{\theta^{4}_{2}(0|\tau)}{\theta^{4}_{3}(0|\tau)} \equiv k^2(e^{i\pi
\tau}), \quad \Im \tau > 0, \ee and that \be \label{kappadef2} 1-\lambda(\tau)
= \frac{\theta^{4}_{4}(0|\tau)}{\theta^{4}_{3}(0|\tau)} \equiv {k'}^2(e^{i\pi
\tau}). \ee The function $\lambda(\tau)$, sometimes also denoted as
$\kappa^2(\tau)$, plays a central role in the theory of modular functions and
modular forms, and a vast literature is devoted to this function - see the
classical monograph \cite{KF}; see also \cite{WW}, \cite{Akh}, \cite{BE},
\cite{math} and Section 3.4 of Chapter 7 in \cite{Ah}. The function
$\lambda(\tau)$ possess several remarkable analytic and arithmetic properties,
some of which are listed in Appendix B.

In terms of the $\lambda$ - modular  function, the formulae
for Renyi read as follows.
\be\label{renlambda1}
   \hskip -2cm
   S_R(\rho_{A},\alpha) = {1 \over 6} \; { \alpha \over 1- \alpha } \; \ln \left( k \; k' \right)
   - {1 \over 12} \; { 1 \over 1-\alpha} \; \ln \Bigl(\lambda(\alpha i \tau_{0})
   (1-\lambda(\alpha i \tau_{0}))\Bigr)
    + {1 \over 3} \ln 2,
\ee
for $h> 2$ and
\be\label{renlambda2}
   \hskip -2cm
   S_R(\rho_{A},\alpha) = {1 \over 6} \; {\alpha \over 1-\alpha } \; \ln
   \left( {k'\over k^2 } \right) + {1 \over 12} \; {1 \over 1-\alpha } \;
   \ln \frac { \lambda^2(\alpha i \tau_{0})}{1-\lambda(\alpha i \tau_{0})}
   + {1 \over 3} \ln 2,
\ee
for $h < 2$. These relations allow to apply to the study of the Renyi
entropy the apparatus of the theory of modular functions. We are going
to address this question specifically in the next publications. Here, we
will only present the two most direct applications of the modular functions
theory related to the symmetry properties of the $\lambda$-function
indicated in (\ref{kappamod1}) - (\ref{landen})).

\subsection{Modular  transformations}

Put
\be \label{fdef}
f(\tau):= \lambda(\tau)(1-\lambda(\tau)),
\quad\mbox{and}\quad
g(\tau) = \frac { \lambda^2(\tau)}{1-\lambda( \tau)},
\ee
and re-write the formulae for the Renyi entropy one more time:
\be\label{renf}
   \hskip -2cm
   S_R(\rho_{A},\alpha) = {1 \over 6} \; { \alpha \over 1- \alpha } \; \ln \left( k \; k' \right)
   - {1 \over 12} \; { 1 \over 1-\alpha} \; \ln f(\alpha i \tau_{0})
    + {1 \over 3} \ln 2
\ee
for $h> 2$, and
\be\label{reng}
   \hskip -2cm
   S_R(\rho_{A},\alpha) = {1 \over 6} \; {\alpha \over 1-\alpha } \; \ln
   \left( {k'\over k^2 } \right) + {1 \over 12} \; {1 \over 1-\alpha } \;
   \ln g(\alpha i \tau_{0})
   + {1 \over 3} \ln 2
\ee
for $h<2$.
The symmetries (\ref{kappamod1}) and (\ref{kappamod2}) imply the
following symmetry relations for $f(\tau)$ and $g(\tau)$ with respect
to the action of the modular group,
\be \label{fmod1}
f(\tau + 1) = -\frac{g(\tau)}{f(\tau)},
\ee
\be \label{fmod2}
f\left(-\frac{1}{\tau}\right) = f(\tau)
\ee
\be \label{gmod1}
g(\tau + 1) = g(\tau),
\ee
\be \label{gmod2}
g\left(-\frac{1}{\tau}\right) = \frac{g(\tau)}{f(\tau)}
\ee
It follows then, that the function $f(\tau)$ is automorphic with respect to the
subgroup of the modular group generated by the transformations,
\be \label{sub2}
\tau \to \tau + 2 \quad \mbox{and}\quad  \tau \to -\frac{1}{\tau},
\ee
while the function $g(\tau)$ is automorphic with respect to the
subgroup of the modular group generated by the transformations,
\be \label{sub3}
\tau \to \tau + 1 \quad \mbox{and}\quad  \tau \to \frac{\tau}{2\tau +1}.
\ee
Of course, the both functions inherit from the lambda-function the
automorphicity with respect to subgroup (\ref{sub1}) (which is a common
subgroup of the subgroups (\ref{sub2}) and (\ref{sub3})).  Therefore,
we arrive at the following conclusion.

{\bf Proposition.} {\it  Up to the trivial addition terms and
multiplicative factors, and after a simple re-scaling, the Renyi
entropy, as a function of $\alpha$, is an automorphic function with
respect to subgroup (\ref{sub2}) of the modular group, in the case
$h>2$, and it is automorphic with respect to  subgroup (\ref{sub3})
of the modular group, in the case $h<2$; in both cases the entropy
is automorphic with respect to subgroup (\ref{sub1}).}

The indicated symmetry  properties of the Renyi entropy yield, in particular,
the following explicit relation between the values of the entropy
at points $\alpha$ and $1/\alpha \tau^2_{0}$.
\be \label{alphasym1}
S_{R}\left(\rho_{A},\frac{1}{\alpha \tau^2_{0}}\right)
=\frac{\alpha \tau^2_{0}}
{\alpha \tau^2_{0} - 1}(1-\alpha)S_{R}(\rho_{A},\alpha)
+\frac{1}{6}\,\frac{1 - \alpha^2 \tau^2_{0}}
{\alpha \tau^2_{0} - 1}\ln\frac{kk'}{4},
\ee
for $h>2$ and
\be \label{alphasym2}
S_{R}\left(\rho_{A},\frac{1}{\alpha \tau^2_{0}}\right)
=\frac{\alpha \tau^2_{0} - \alpha^2 \tau^2_{0}}
{\alpha \tau^2_{0} - 1}S_{R}(\rho_{A},\alpha)
+\frac{1}{6}\,\frac{1 - \alpha^2 \tau^2_{0}}
{\alpha \tau^2_{0} - 1}\ln\frac{k'}{4k^2}
\ee
$$
 - \frac{1}{12}
\frac{\alpha \tau^2_{0}}
{\alpha \tau^2_{0} - 1}\ln f(\alpha i \tau_{0}),
$$
for $h<2$. We  bring the attention of the reader to the appearance
in the case $h<2$ of an extra term involving the modular function
$f(\tau)$ .

\subsection{$\alpha = 2^n$}
For the indicated values of the parameter $\alpha$ one can apply
Landen's transformation (\ref{landen}) and
reduce the function $\lambda(\alpha i \tau_{0})$
to the function
$$
\lambda(i\tau_{0}) \equiv k^2.
$$
Hence, for these values of $\alpha$ the Renyi entropy becomes an
{\it elementary} function of the initial physical parameters $h$ and
$\gamma$. Let us demonstrate this in the case $\alpha = 2$.

From (\ref{landen}) it follows that
$$
\lambda(2i\tau_{0}) = \left(\frac{1-k'}{1+k'}\right)^2.
$$
Therefore,
$$
f(2i\tau_{0}) = \frac{4k'(1-k')^2}{(1+k')^4}\quad
\mbox{and}\quad
g(2i\tau_{0})= \frac{(1-k')^4}{4k'(1+k')^2}.
$$
Using these, we can find the Renyi entropy
for $\alpha = 2$ from (\ref{renf}) and (\ref{reng}):
\be
   S_R (\alpha=2) = -{1 \over 6} \; \ln
   \left( k^2 \; k'^{3/2} \; { (1 + k')^2 \over (1-k') } \right) + {1 \over 2} \ln 2,
\ee
for $h>2$ and
\be
   \hskip -1cm
   S_R (\alpha = 2 ) = -{1 \over 6} \; \ln
   \left( {k'^{3/2} \over k^4 } \; { (1-k')^2 \over (1+k') } \right)
   + {1 \over 2} \ln 2,
\ee
for $h<2$. Repeating Landen's transformation again and again,
we can iteratively construct a ladder of ``elementary'' entropies
for increasing values of $\alpha = 2^n$.

%\section{A second, final remark}

%As Peschel noted in his paper, the formulae for the Von Neumann
%entropy (\ref{A1},\ref{A2}) can be written as
%\bea
%   S & = & 2 \sum_j \ln \left( 1 + \eu^{- j \epsilon} \right) +
%   2 \sum_j j \epsilon \; {\eu^{-j \epsilon} \over 1 + \eu^{-j \epsilon} }
%   \nonumber \\
%   & = & \ln Z + U ,
%\eea where $j = 2m$ or $j=2m+1$ depending on the case considered.

%Now, let me generalize this formula by introducing a parameter
%$\alpha$ as
%\bea
%   S_\alpha & = & 2 \sum_j \ln \left( 1 + \eu^{- j \; \alpha \; \epsilon} \right) +
%   2 \sum_j j \epsilon \; {\eu^{-j \; \alpha \; \epsilon} \over 1 + \eu^{-j \; \alpha \; \epsilon} }
%   \nonumber \\
%   & = & \ln Z_\alpha + U_\alpha .
%\eea Note that $\ln Z_\alpha$ is what appears in the expression
%for Renyi entropy.

%The remarkable fact I noticed is
%\bea
%   - {\de \over \de \alpha} \ln Z_\alpha & = &
%   - {\de \over \de \alpha} 2 \sum_j \ln \left( 1 + \eu^{- j \; \alpha \; \epsilon} \right)
%   \nonumber \\
%   & = & 2 \sum_j j \; \epsilon \; { \eu^{- j \; \alpha \; \epsilon} \over 1 + \eu^{- j \; \alpha \; \epsilon} }
%   = U_\alpha.
%\eea This means that the Von Neumann entropy could be written as
%\be
%   S = \left. \left( 1 - {\de \over \de \alpha} \right) \ln Z_\alpha \right|_{\alpha = 1}
%\ee and generates interesting relations between the elliptic
%functions.

%I don't know yet whether this remark means something or whether it
%can help in the calculation of Renyi entropy. Maybe you can tell
%me if this makes any bell ring.

\section{Summary and Conclusions}

We analyzed the entanglement of the ground state of the infinite
one-dimensional $XY$ spin chain by calculating the Renyi entropy $S_\alpha
(\rho_A)$ of a large block $A$ of neighboring spins. The Renyi entropy has been
proposed as a meaningful measure of the quantum entanglement of a system and it
is a natural generalization of the Von Neumann entropy. In fact, for $\alpha
=1$ the quantities are equal. Moreover, knowledge of the Renyi entropy for all
$\alpha$'s allow for the reconstruction of the density matrix and an easier
identification of the sources of entanglement in the mixed quantum state.

We arrived at an analytic expression of the entropy in the bulk of the
two-dimensional phase diagram of the model, in terms of an elliptic parameter
and elliptic theta functions. These expressions allowed us to study the
behavior of the Renyi entropy for the different values of $\alpha$ and of the
parameters of the model. We found the limiting behavior of the entropy for
$\alpha \to \infty$, which is essentially the {\it single copy entanglement}
introduced in \cite{EisCram}. In that work, it was shown that this quantity
scales like $1/6 \ln L$ for the isotropic $XX$ model. This is consistent with
our findings -- setting $k \sim 1$, $k' \sim 0$ in  (\ref{largealpha2})-- and
we generalize it to the rest of the phase diagram.

In the limit $\alpha \to 0$ we showed that the entropy diverges like
$\alpha^{-1}$. A very interesting behavior occurs at the factorizing field $h_f
(\gamma) = 2 \sqrt{ 1 - \gamma^2}$. On this line the ground state can be
written as a sum of two product states. This means that the reduced density
matrix remains proportional to the two-dimensional identity matrix and we
showed that the Renyi entropy is $S_\alpha = \ln 2$, independent from $\alpha$.
So, even for $\alpha \to 0$ the Renyi entropy stays finite at the factorizing
field, while it diverges as one moves away from this line.

The bulk of the $XY$ model is gapped and the entropy of a large block is known
to saturate to a finite value, which we calculated. As one approaches the
critical lines, the entropy diverges logarithmically in the gap size. We
calculated exactly the prefactor of this logarithmic divergence as a function
of $\alpha$ for the two universality classes of the critical lines and found
agreement with the Von Neumann result at $\alpha = 1$, as to be expected.

Finally, using the properties of the theta functions, we showed that the
limiting Renyi entropy is a modular function of $\alpha$. The properties of the
entropy under modular transformations seem very interesting and will be the
subject of a subsequent paper. In a previous work \cite{fijk2} we showed that
the curves of constant entropy are ellipses and hyperbolae and that they all
meet at the point $(h,\gamma)=(2,0)$, which is a point of high singularity for
the entropy. This is valid also for the Renyi entropy and seems to be connected
with the aforementioned modular properties of the entropy. We will investigate
this relationship in a future work.

\vskip 0.5cm

\section*{Acknowledgments}

We are grateful to Dr. Bai Qi Jin for his work on the analytical properties of
the Renyi entropy about the variable $\alpha$, as it appears in equations
(\ref{ren_neu1}) and (\ref{ren_neu2}). F.F. would like to thank Alexander
Abanov, Siddhartha Lal and most of all Giuseppe Mussardo for their help and
availability for discussions. This work has been partially supported by the NFS
grant DMS-0503712 (V.E.K.), DMS-0401009 and DMS-0701768 (A.R.I.).

\vskip 0.5cm

\appendix

%\section*{Appendix}

\section{Theta Functions}

In this appendix the necessary facts of the theory of Jacobi
theta-functions are presented. For more detail, we refer the reader
to any standard text book on elliptic functions, e.g.\cite{WW}.

Among the four theta-functions, only one is functionally
independent, and usually it is taken to be the function
$\theta_{3}(z|\tau)$. The rest of the theta-functions are related to
$\theta_{3}(z|\tau)$ via the simple equations, \be \label{theta13}
\theta_{1}(z| \tau) = -ie^{\frac{\pi i \tau}{4} + iz}
\theta_{3}\left(z + \frac{1}{2}\,\pi + \frac{1}{2}\,\pi\tau\,|
\tau\right), \ee \be \label{theta23} \theta_{2}(z| \tau) =
e^{\frac{\pi i \tau}{4} + iz} \theta_{3}\left(z +
\frac{1}{2}\,\pi\tau\,| \tau\right), \ee \be \label{theta43}
\theta_{4}(z| \tau) = \theta_{3}\left(z + \frac{1}{2}\,\pi\,|
\tau\right), \ee

The principal characteristic properties of the theta-functions
are their quasi -  periodicity properties with respect to
the shifts, $z \to z +\pi$ and $z \to z +\pi \tau$:
\be \label{thperiod11}
\theta_{1}(z+\pi|\tau) = -\theta_{1}(z|\tau),
\ee
\be \label{thperiod12}
\theta_{1}(z + \pi \tau|\tau)
=-e^{-\pi i \tau -2iz}\theta_{1}(z |\tau),
\ee
\be \label{thperiod21}
\theta_{2}(z+\pi|\tau) = -\theta_{2}(z|\tau),
\ee
\be \label{thperiod22}
\theta_{2}(z + \pi \tau|\tau)
=e^{-\pi i \tau -2iz}\theta_{2}(z |\tau),
\ee
\be \label{thperiod31}
\theta_{3}(z+\pi|\tau) = \theta_{3}(z|\tau),
\ee
\be \label{thperiod32}
\theta_{3}(z + \pi \tau|\tau)
= e^{-\pi i \tau -2iz}\theta_{3}(z|\tau),
\ee
\be \label{thperiod41}
\theta_{4}(z+\pi|\tau) = \theta_{4}(z|\tau),
\ee
\be \label{thperiod42}
\theta_{4}(z + \pi \tau|\tau)
=-e^{-\pi i \tau -2iz}\theta_{4}(z|\tau).
\ee
The complementary set of the properties is the set of the following
symmetry relations with respect
to the transformations, $\tau \to \tau + 1$ and $\tau \to -\tau^{-1}$
(that is, with respect to the action of the modular group):
\be \label{thmod11}
\theta_{1}(z|\tau+1) = e^{\frac{\pi i}{4}}\theta_{1}(z|\tau),
\ee
\be \label{thmod12}
\theta_{1}\left(\frac{z}{\tau}\,|-\frac{1}{\tau}\right)
=\frac{1}{i}\,\sqrt{\frac{\tau}{i}}\,e^{\frac{iz^2}{\pi \tau}}\,
\theta_{1}(z|\tau),
\ee
\be \label{thmod21}
\theta_{2}(z|\tau+1) = e^{\frac{\pi i}{4}}\theta_{2}(z|\tau),
\ee
\be \label{thmod22}
\theta_{2}\left(\frac{z}{\tau}\,|-\frac{1}{\tau}\right)
=\sqrt{\frac{\tau}{i}}\,e^{\frac{iz^2}{\pi \tau}}\,
\theta_{4}(z|\tau),
\ee
\be \label{thmod31}
\theta_{3}(z|\tau+1) = \theta_{4}(z|\tau),
\ee
\be \label{thmod32}
\theta_{3}\left(\frac{z}{\tau}\,|-\frac{1}{\tau}\right)
=\sqrt{\frac{\tau}{i}}\,e^{\frac{iz^2}{\pi \tau}}\,
\theta_{3}(z|\tau),
\ee
\be \label{thmod41}
\theta_{4}(z|\tau+1) = \theta_{3}(z|\tau),
\ee
\be \label{thmod42}
\theta_{4}\left(\frac{z}{\tau}\,|-\frac{1}{\tau}\right)
=\sqrt{\frac{\tau}{i}}\,e^{\frac{iz^2}{\pi \tau}}\,
\theta_{2}(z|\tau),
\ee
where the branch of the square root is fixed by the condition,
$$
\sqrt{\frac{\tau}{i}} = 1,\quad \mbox{if}\quad \tau =i.
$$
An immediate important corollary of these relations is the
possibility of the following alternative series representations
(the Jacobi identities) for
the theta-functions participating in the formulae (\ref{renfinal2new})
and (\ref{renfinal4new}) for the Renyi entropy.

\be \label{jacobi2}
\theta_{2}(0|\tau) = 2\sum_{n=0}^{\infty}e^{\pi i\tau \left(n+\frac{1}{2}\right)^2}
= \sqrt{\frac{i}{\tau}}\left(
1 + 2 \sum_{n=1}^{\infty}(-1)^ne^{-\frac{\pi i n^2}{\tau}}\right),
\ee

\be \label{jacobi3}
\theta_{3}(0|\tau) =1 + 2\sum_{n=1}^{\infty}e^{\pi i\tau n^2}
= \sqrt{\frac{i}{\tau}}\left(
1 + 2 \sum_{n=1}^{\infty}e^{-\frac{\pi i n^2}{\tau}}\right),
\ee

\be \label{jacobi4} \theta_{4}(0|\tau) =1 + 2\sum_{n=1}^{\infty}(-1)^ne^{\pi
i\tau n^2} = 2\sqrt{\frac{i}{\tau}}\sum_{n=0}^{\infty} e^{-\frac{\pi
i}{\tau}\left(n+\frac{1}{2}\right)^2}. \ee The first series in each of these
formulae allow an efficient evaluation of the corresponding theta-constant for
large $\Im \tau$, while the second series provides a tool for analysis  of the
theta-constant in the limit of small $|\tau|$. In Section 4 we  use these
identities for investigating the singularity of the Renyi entropy at $\alpha
=0$.

The last general fact of the theory of elliptic theta-function we will need, is
the description of their zeros, as the functions of the first argument. In view
of the relations (\ref{theta13}) - (\ref{theta43}) it is sufficient to describe
the zeros of $\theta_{3}(z|\tau)$. They are:
$$
z \equiv z_{nm} =
\frac{1}{2}\,\pi + \frac{1}{2}\,\pi \tau + n\pi + m\pi\tau, \quad n, m(\ref{renfinal2new}) \in {\Bbb Z}.
$$
This information about the zeros of $\theta_{3}(z|\tau)$,  taking in
conjunction with the relations (\ref{theta23}) and (\ref{theta43}), implies, in
particular, that \be \label{nonzero}
\theta_{2}(0|\tau)\theta_{3}(0|\tau)\theta_{4}(0|\tau)\neq 0,\quad \forall
\tau, \,\, \Im \tau > 0. \ee

\section{Elliptic Lambda Function}

The properties of the $\lambda$-function presented below form an
important but very far from being exhausted set of the extremely
exciting properties and connections which this function enjoys. For
more on the lambda and related functions we refer the reader, in
addition to the monographs already mentioned, to the websites
\cite{Weis1} and \cite{Weis2} and to the references and links
indicated there.

\begin{enumerate}
\item Let $\Omega$ denote the ``triangle'' on the Lobachevsky upper half
$\tau$ - plane with the vertices at the points $0$, $1$ and $\infty$ and
with the zero angle at each vertex (the edges are: $\Re \tau = 0$,
$\Re \tau = 1$, $|\tau -1/2| = 1/2$). Then, the function $w = \lambda(\tau)$
performs the conformal mapping of the triangle $\Omega$ onto the upper-half plane
$\Im w > 0$, and it sends the vertices  $0$, $1$, and $\infty$ to the
points $1$, $\infty$, and $0$, respectively. It also should be
noticed that the real line is the {\it natural boundary} for $\lambda(\tau)$ -
the function can not be analytically  continued beyond it.
\item The direct corollary of the conformal property just stated is the
following analytic fact. Let $\{f,z\}$ denotes the Schwartz derivative,
$$
\{f,z\} = \frac{f'''(z)}{f'(z)} - \frac{3}{2}\left(\frac{f'''(z)}{f'(z)}\right)^{2}.
$$
Then, the lambda-function $\lambda(\tau)$ satisfies the following
differential equation,
\be \label{lambdadif}
\{\lambda,\tau\}
= -\frac{1}{2}\frac{1}{\lambda^{2}} -\frac{1}{2}\frac{1}{(\lambda - 1)^2}
+\frac{1}{\lambda(\lambda - 1)}
\ee
\item The function $\lambda(\tau)$ satisfies the following symmetry
relations with respect to the actions of the generators of the
modular group (cf. (\ref{thmod21}) - (\ref{thmod42})),
\be \label{kappamod1}
\lambda(\tau + 1) = \frac{\lambda(\tau)}{\lambda(\tau) - 1},
\ee
\be \label{kappamod2}
\lambda\left(-\frac{1}{\tau}\right) = 1 - \lambda(\tau).
\ee
These symmetries in turn imply the equations,
\be \label{kappamod3}
\lambda(\tau + 2) = \lambda(\tau),
\ee
\be \label{kappamod4}
\lambda\left(\frac{\tau}{2\tau+1}\right) = \lambda(\tau),
\ee
which show that the function $\lambda(\tau)$ is automorphic
function with respect to the subgroup of the modular group generated
by the transformations,
\be \label{sub1}
\tau \to \tau + 2, \quad \mbox{and}\quad \tau \to \frac{\tau}{2\tau+1}.
\ee
\item In addition to the symmetries with respect to the modular
group, the function $\lambda(\tau)$ satisfies the so-called second
order transformation, also called {\it  Landen's transformation}, which
describes the action on $\lambda(\tau)$ of the doubling  map, $\tau \to 2\tau$:
\be \label{landen}
\sqrt{\lambda(2\tau)} = \frac{1-\sqrt{1-\lambda(\tau)}}{1+\sqrt{1-\lambda(\tau)}}.
\ee
Here, the branches of the square roots are chosen according
to the equations (cf. (\ref{kappadef}) and (\ref{kappadef2})),
$$
\sqrt{\lambda(\tau)} = \frac{\theta^{2}_{2}(0|\tau)}{\theta^{2}_{3}(0|\tau)}
\equiv k(e^{i\pi\tau}),
\quad
\sqrt{1-\lambda(\tau)} = \frac{\theta^{2}_{4}(0|\tau)}{\theta^{2}_{3}(0|\tau)}
\equiv k'(e^{i\pi\tau}).
$$
\item By means of the algebraic equation,
\be \label{Jtau}
J(\tau) = \frac{4}{27}\frac{(1-\lambda(\tau) + \lambda^2(\tau))^3}{\lambda^2(\tau)
(1-\lambda(\tau))^2},
\ee
the elliptic lambda - function defines even more fundamental object of the theory
of modular forms -
{\it Klein's absolute invariant}  $J(\tau)$. The function $J(\tau)$ is a {\it modular function},
i.e. it is automorphic with respect to the modular group itself,
\be \label{Jmod}
J(\tau + 1) = J(\tau), \quad J\left(-\frac{1}{\tau}\right) = J(\tau);
\ee
moreover, any other modular function is algebraically expressible in terms
of the invariant $J(\tau)$.  The function $J(\tau)$ admits also an alternative representation
in terms of the Ramanujan-Eisenstein series $E_{j}$:
\be \label{JEis}
J(\tau) =   \frac{E^{3}_4(\tau)}{E^{3}_4(\tau)-E^{2}_6(\tau)}.
\ee
We remind that
$$
E_4(\tau)   =   1+240\sum_{k=1}^{\infty}\sigma_3(k)q^{2k},
\quad   E_6(\tau)   =   1-504\sum_{k=1}^{\infty}\sigma_5(k)q^{2k},
\,\,\,\, q = e^{i\pi \tau},
$$
where $\sigma_k(n)$ is a divisor function, i.e.
$$
\sigma_k(n)=\sum_{d|n}d^k.
$$
\end{enumerate}


\begin{thebibliography}{99}

\bibitem{ben}{C.H.\ Bennett, H.J.\ Bernstein, S.\ Popescu, and B.\ Schumacher, Phys. Rev. {\bf A 53}, 2046, (1996)}

\bibitem{vidal}{G.\ Vidal, J.I.\ Latorre, E.\ Rico, and A.\ Kitaev,
    Phys. Rev. Lett. {\bf 90}, 227902, (2003)}

\bibitem{vidal1}{J.I.\ Latorre, E.\ Rico, and G.\ Vidal, arXiv:
    quant-ph/0304098}

\bibitem{cardy} Calabrese  P and Cardy J 2004  {\it J. Stat. Mech.: Theor.
    Exp.} P0406002

\bibitem{GRAC} Vedral V 2003 {\it Nature} {\bf 425} 28; Ghosh S, Rosenbaum T F,
    Aeppli G and Coppersmith S N 2003 {\it Nature} {\bf 425} 48

\bibitem{keat} Keating J P and Mezzadri F {\it Preprint} quant-ph/0407047

\bibitem{pes} Peschel I Journal of Statistical Mechanics (2004) P12005

\bibitem{renyi}{A.\ R\'enyi, {\it Probability Theory}, North-Holland,
    Amsterdam, 1970 }

\bibitem{abe}{S.\ Abe and A.\ K.\ Rajagopal, Phys. Rev. {\bf A 60}, 3461,
  (1999)}

\bibitem{BD} Bennett C H and DiVincenzo D P 2000 {\it Nature} {\bf 404} 247

\bibitem{hb} H. E. Brandt, {\it  Quantum Information and Computation IV, Proc.
    SPIE},Vol. 6244, Bellingham, Washington (2006) pp. 62440G-1-8.

\bibitem{L}Lloyd S 1993 {\it Science} {\bf 261} 1569;  1994 {\cal ibid} {\bf
    263} 695

\bibitem{Lieb}{E.\ Lieb, T.\ Schultz and D.\ Mattis, Ann. Phys. {\bf 16}, 407,
    (1961)}

\bibitem{mccoy}{E.\ Barouch and B.M.\ McCoy, Phys. Rev. {\bf A 3}, 786, (1971)}

\bibitem{mccoy2}{E.\ Barouch, B.M.\ McCoy and M.\ Dresden, Phys. Rev. {\bf A 2}, 1075, (1970)}

\bibitem{gallavotti}{D.B.\ Abraham, E.\ Barouch, G.\ Gallavotti and A.\ Martin-L\"of,
    Phys. Rev. Lett. {\bf 25}, 1449, (1970); Studies in Appl. Math. {\bf 50}, 121, (1971); {\cal ibid} {\bf 51}, 211, (1972)}

\bibitem{shrock}
    G. M\"uller, and R.E. Shrock, Phys. Rev. {\bf B 32}, 5845 (1985).
    J. Kurmann, H. Thomas, and G. M\"uller, Physica {\bf A 112}, 235 (1982);

\bibitem{arealaw} K. Audenaert, J. Eisert, M.B. Plenio, R.F. Werner, {\it Phys.
    Rev. } {\bf A 66}, 042327 (2002);
    Norbert Schuch, Michael M. Wolf, Frank Verstraete, J. Ignacio Cirac,
    arXiv:0705.0292; M. B. Hastings, {\it JSTAT}, P08024 (2007).

\bibitem{jin} Jin B Q and Korepin V E 2004 {\it J. Stat. Phys.} {\bf 116} 79

\bibitem{ijk1}A. R. Its, B.-Q. Jin, V. E. Korepin, Journal Phys. A: Math. Gen.
    vol 38, pages 2975-2990, 2005, quant-ph/0409027

\bibitem{ijk2}A. R. Its, B.-Q. Jin, V. E. Korepin,quant-ph/0606178

\bibitem{fijk1}F. Franchini, A. R. Its, B.-Q. Jin, V. E. Korepin,
    quant-ph/0606240

\bibitem{fijk2}F. Franchini, A. R. Its, B.-Q. Jin, V. E. Korepin, {\it J.
    Phys.} {\bf A 40} (2007) 8467-8478

\bibitem{wu}{T.T.\ Wu, Phys. Rev. {\bf 149}, 380, (1966)}

\bibitem{fisher}{M.E.\ Fisher and R.E.\ Hartwig, Adv. Chem. Phys. {\bf 15},
    333, (1968)}

\bibitem{basor}{E.L.\ Basor, Indiana Math. J. {\bf 28}, 975, (1979)}

\bibitem{tracy}{E.L.\ Basor and C.A.\ Tracy, Physica {\bf A 177}, 167, (1991)}

\bibitem{abbs}{A.\ B\"ottcher and B.\ Silbermann, {\it Analysis of
      Toeplitz Operators}, Springer-Verlag, Berlin, 1990}

\bibitem{aban} Shiroishi M, Takahahsi M and Nishiyama Y 2001 {\it J. Phys. Soc.
    Jpn} {\bf 70} 3535; Abanov A G and Franchini F 2003 {\it Phys. Lett.}
    A{\bf 316} 342

\bibitem{iiks} Its A R, Izergin A G, Korepin V E and Slavnov N A 1990 {\it Int.
    J. Mod. Phys.} B {\bf 4} 1003;

\bibitem{sla} Its A R, Izergin A G, Korepin V E and Slavnov N A 1993 {\it Phys.
    Rev. Lett.} {\bf 70 } 1704

\bibitem{korepin}{N.M.\ Bogoliubov, A.G.\ Izergin, and V.E.\ Korepin,
    {\it Quantum Inverse Scattering Method and Correlation Functions},
    Cambridge Univ. Press, Cambridge, 1993}

\bibitem{jpaits}{A. R. Its, B.-Q. Jin and V.E. Korepin, J. Phys. {A 38}, 2975,
    (2005)}

%\bibitem{tsallis}{C. Tsallis, J. Stat. Phys. {\bf 52}, 479, (1988)}

\bibitem{WW} E. T. Whittaker and G. N. Watson, {\it A Course of Modern
    Analysis}, Cambridge at the University Press 1927

\bibitem{EisCram} J. Eisert and M. Cramer, {\it Phys. Rev.} {\bf A 72}, 042112
    (2005).

\bibitem{KF} F. Klein, R. Fricke, {\it Vorlesungen\"uber die Theorie der
    elliptischen Modulfunktionen}, vol. 2, B. G. Teubner, Leipzig, 1890-1892.

\bibitem{Akh} N. I. Akhiezer, {\it Elements of the Theory of Elliptic
    Functions}, Translations of Mathematical Monographs, volume 79, AMS, 1990

\bibitem {BE} H. Bateman, A. Erdelyi, {\it Higher Transcendental Functions},
    McGraw-Hill, NY, 1953.

\bibitem{math}{M. Abramowitz and I. Stegun, (eds.) Handbook of Mathematical
    Functions, Dover, New York (1965)}

\bibitem{Ah} L. Ahlfors, {\it Complex Analysis}, 3d edition, McGraw-Hill, Inc,
    1979.

\bibitem{Weis1} Weisstein, Eric W. "Elliptic Lambda Function." From
    MathWorld--A Wolfram Web Resource.
    http://mathworld.wolfram.com/EllipticLambdaFunction.html

\bibitem{Weis2} Weisstein, Eric W. "Klein's Absolute Invariant." From
    MathWorld--A Wolfram Web Resource.
    http://mathworld.wolfram.com/KleinsAbsoluteInvariant.html

%\bibitem{Henley}{C.A.\ Cheong and C.L.\ Henley, arXiv:cond-mat/0206196}

\end{thebibliography}
\end{document}